\documentclass[showpacs,twocolumn,showkeys,amsmath,amssymb,pra]{revtex4-1}

\usepackage{bm}
\usepackage{color}
\usepackage{graphicx}
\newcommand{\rd}{{\mathrm d}}
\newcommand{\re}{{\mathrm e}}
\newcommand{\ri}{{\mathrm i}}
\newcommand{\lF}{{\langle\!\langle}}
\newcommand{\rF}{{\rangle\!\rangle}}
\newcommand{\ab}{a^{\phantom{\dagger}}}
\newcommand{\ad}{a^{\dagger}}
\begin{document}

\title[Floquet condensates]
      {Adiabatic preparation of Floquet condensates}

\author{Christoph Heinisch}
\author{Martin Holthaus}

\affiliation{Institut f\"ur Physik, Carl von Ossietzky Universit\"at, 
	D-26111 Oldenburg, Germany}
                  
\date{December 24, 2015}

\begin{abstract}
We argue that a Bose-Einstein condensate can be transformed into a Floquet 
condensate, that is, into a periodically time-dependent many-particle state 
possessing the coherence properties of a mesoscopically occupied 
single-particle Floquet state. Our reasoning is based on the observation 
that the denseness of the many-body system's quasienergy spectrum does not 
necessarily obstruct effectively adiabatic transport. Employing the idealized 
model of a driven bosonic Josephson junction, we demonstrate that only a small 
amount of Floquet entropy is generated when a driving force with judiciously
chosen frequency and maximum amplitude is turned on smoothly.
\end{abstract} 




\maketitle 


\section{Introduction}
\label{sec:1}

The study of ultracold atoms in optical lattices under the influence of 
time-periodic external forcing has gained tremendous momentum recently. 
Activities in this field include the realization of tunable gauge 
potentials~\cite{StruckEtAl12} and topological insulators~\cite{HaukeEtAl12} 
with ultracold atoms in periodically shaken optical lattices~\cite{Eckardt15}, 
the simulation of effective ferromagnetic domains~\cite{ParkerEtAl13} and of 
the roton-maxon dispersion known from superfluid helium~\cite{HaEtAl15}, 
the realization of the topological Haldane model with ultracold 
fermions~\cite{JotzuEtAl14}, the observation of Bose-Einstein condensation 
in strong synthetic magnetic fields~\cite{KennedyEtAl15}, and the detection 
of multiphoton-like transitions with quantum gases in driven optical 
lattices~\cite{WeinbergEtAl15}. Further theoretical proposals have addressed
the creation of Majorana fermions~\cite{JiangEtAl13,LiuEtAl13} and of 
topologically protected edge states~\cite{ReichlMueller14} in driven cold-atom 
quantum systems. The diversity of this list, which is still far from complete, 
suggests that the addition of time-periodic forcing to the toolbox of 
ultracold-atoms physics may well constitute a decisive step towards efficient 
quantum simulation.   

The common theme underlying this development is the use of the Floquet
picture for periodically time-dependent quantum systems~\cite{Holthaus15}.
The formal content of this approach is easy to formulate: Consider a quantum
system defined on some Hilbert space ${\mathcal H}$, be it a single-particle 
or a many-body system, the dynamics of which are governed by an explicitly
time-dependent Hamiltonian $H(t)$ which is periodic in time with period~$T$,  
\begin{equation}
	H(t) = H(t+T) \; .
\label{eq:HOT}	
\end{equation}
Then Floquet's theorem~\cite{Floquet83} suggests the existence of a set of 
particular solutions to the time-dependent Schr\"odinger equation possessing
the form   
\begin{equation}
	| \psi_n(t) \rangle = | u_n(t) \rangle \exp(-\ri\varepsilon_n t/\hbar)
	\; , 
\label{eq:FLS}
\end{equation}	
where the Floquet functions $| u_n(t) \rangle$ inherit the $T$-periodicity
of the Hamiltonian, so that 
\begin{equation}
	| u_n(t) \rangle =  | u_n(t+T) \rangle \; ;  
\end{equation}	
the phase factors accompanying their time-evolution are determined by the
quasienergies~$\varepsilon_n$~\cite{Zeldovich66,Ritus66}. Actually the 
existence of square-integrable Floquet states~(\ref{eq:FLS}) is subject 
to severe mathematical complications in the general case of systems possessing 
an infinite-dimensional Hilbert space~\cite{Howland92}. Fortunately, in many 
cases of practical interest the dynamics remain confined to an effectively 
finite-dimensional~${\mathcal H}$, so that these complications may be 
neglected. Then the Floquet states form a complete system in ${\mathcal H}$ 
at each instant~$t$, and {\em every\/} solution $|\psi(t) \rangle$ to the 
time-dependent Schr\"odinger equation can be expanded with respect to this
basis,    
\begin{equation}
 	|\psi(t) \rangle =  \sum_n a_n | u_n(t) \rangle 
	\exp(-\ri\varepsilon_n t/\hbar) \; ,	
\label{eq:EXP}
\end{equation}
with coefficients $a_n$ remaining constant in time, since the periodic
time-dependence of the Hamiltonian has already been incorporated into the
basis states themselves. In other words, under conditions of perfect 
isolation, guaranteeing unitary evolution, the Floquet states~(\ref{eq:FLS}) 
are equipped with constant occupation probabilities $|a_n|^2$.

However, in an actual experiment the time-periodic influence will not last
forever. Rather, it has to be turned on at some point, and will be turned off 
later. That is, instead of a perfectly time-periodic Hamiltonian~(\ref{eq:HOT})  
one is more likely to encounter a Hamiltonian of the form
\begin{equation}
	H(t) = H_0 +  H_1(t) \; ,
\end{equation}	
where $H_0$ describes the isolated, ultracold quantum gas as long as it is 
left to itself, while $H_1(t)$ models an external force that is initially
absent, then switched on in some way or other, is time-periodic only for
a finite number of periods~$T$, and is finally switched off. In such cases
the expansion~(\ref{eq:EXP}) holds in the middle interval, after the periodic
forcing has been turned on and before it is turned off again, which may
coincide with the interval during which the measurements are performed. But 
then the precise manner in which the forcing has been turned on is of crucial 
importance for the entire experiment: Unless there is a further relaxation 
mechanism, the preserved occupation amplitudes $a_n$ of the individual Floquet 
states during the action of the periodic force are determined solely by its 
turn-on.

This observation sets the stage for the current paper. We will demonstrate
that it is possible to prepare a {\em  Floquet condensate\/}, that is, 
a bosonic many-body Floquet state possessing the coherence properties 
of a macroscopically occupied single-particle Floquet 
state~\cite{GertjerenkenHolthaus14a,GertjerenkenHolthaus14b}, if the external
force is switched on in an effectively adiabatic manner, provided the forcing 
strength does not exceed a critical value which depends on the number of
particles. Although we employ the idealized model of a periodically driven 
bosonic Josephson junction~\cite{HolthausStenholm01} for numerical 
demonstration purposes, the main qualitative features derived from that model 
may be valid in more general settings.
   
We proceed as follows: In Sec.~\ref{sec:2} we recapitulate the adiabatic
principle for Floquet states~\cite{BreuerHolthaus89a,BreuerHolthaus89b,
DreseHolthaus99}, and show that the adiabatic transport of Floquet states is 
accompanied by a Berry phase. In Sec.~\ref{sec:3} we then discuss numerical 
model calculations which clarify how this principle works in practice in 
a bosonic many-body system, and elucidate why it has a limited regime of 
applicability under conditions of rather strong forcing. In the final 
Sec.~\ref{sec:4} we then formulate our conclusions, aiming at model-independent
predictions.

\section{The adiabatic principle for Floquet states}
\label{sec:2}

Let us assume that the Hamiltonian of an externally forced ultracold-atoms
system depends on a set of slowly changing parameters
\begin{equation}
	\bm P (t) = \big( P_1(t), P_2(t), \ldots \big) \; ,
\end{equation}
such that it is strictly periodic in time when these parameters are kept 
{\em fixed\/} at instantaneous values~$\bm P$,  
\begin{equation}
	H^{\bm P}(t) = 	H^{\bm P}(t+T) \; .
\label{eq:HPT}	
\end{equation}	
For instance, $P_1(t)$ may denote the slowly changing envelope of a sinusoidal 
force with angular frequency $\omega = 2\pi/T$, as in the example considered 
later in Sec.~\ref{sec:3}; the term ``slow'' then means ``slow compared to the 
cycle time~$T$''. In principle, also the driving frequency $\omega$ may be 
varied in an adiabatic manner~\cite{DreseHolthaus99}. The task now is to solve 
the time-dependent Schr\"odinger equation with moving parameters, 
\begin{equation}
	\ri\hbar\frac{\rd}{\rd t} | \psi(t) \rangle =
	H^{\bm P (t)}(t) | \psi(t) \rangle \; .
\label{eq:SGL}
\end{equation}
Because this parameter motion is supposed to occur slowly, it is useful to 
invoke the {\em instantaneous Floquet states\/}  
\begin{equation}
 	| \psi_n^{\bm P}(t) \rangle = | u_n^{\bm P}(t) \rangle 
	\exp(-\ri\varepsilon_n^{\bm P} t/\hbar)
\label{eq:IFS}		
\end{equation}
associated with the Hamiltonian operators~(\ref{eq:HPT}) for each fixed 
set~$\bm P$ encountered in the course of time. These states~(\ref{eq:IFS}) 
obviously obey the equation
\begin{eqnarray} & &
	\ri\hbar \frac{\rd}{\rd t} | \psi_n^{\bm P}(t) \rangle 
\nonumber \\	& = &
	\left( \ri\hbar \frac{\rd}{\rd t} | u_n^{\bm P}(t) \rangle
	+ \varepsilon_n^{\bm P} | u_n^{\bm P}(t) \rangle \right) 
	\exp(-\ri\varepsilon_n^{\bm P} t/\hbar)
\nonumber \\	& = &
	H^{\bm P}(t) | u_n^{\bm P}(t) \rangle
	\exp(-\ri\varepsilon_n^{\bm P} t/\hbar) \; ,
	\phantom{\frac{\rd}{\rd t}}
\end{eqnarray}		
giving
\begin{equation}
	\left( H^{\bm P}(t) - \ri\hbar \frac{\rd}{\rd t} \right)
	| u_n^{\bm P}(t) \rangle = 
	\varepsilon_n^{\bm P} | u_n^{\bm P}(t) \rangle \; .
\label{eq:EVP}
\end{equation}
This is an {\em eigenvalue equation\/} for the time-periodic Floquet functions
$| u_n^{\bm P}(t) \rangle$, providing the quasienergies $\varepsilon_n^{\bm P}$
as their eigenvalues, quite similar to a stationary Schr\"odinger equation 
which yields the energy eigenvalues and eigenfunctions of a time-independent 
Hamiltonian. However, this eigenvalue problem~(\ref{eq:EVP}) is no longer 
posed in the system's physical Hilbert space ${\mathcal H}$, because one has 
to incorporate the periodic boundary conditions
\begin{equation}
	| u_n^{\bm P}(t) \rangle = | u_n^{\bm P}(t+T) \rangle \; .
\label{eq:PBC}
\end{equation}
To this end, one introduces the {\em extended Hilbert space\/}  
$L_2[0,T] \otimes {\mathcal H}$, consisting of $T$-periodic square-integrable 
functions, in which the time~$t$ is treated on the same footing as the spatial 
coordinates. Accordingly, the scalar product in this extended space is given 
by~\cite{Sambe73}
\begin{equation}
	\lF u | v \rF = \frac{1}{T}\int_0^T \! \rd t \, 
	\langle u(t) | v(t) \rangle \; ,
\end{equation}	
naturally involving integration over the ``time coordinate''. Following 
Sambe~\cite{Sambe73}, one writes $| u_n^{\bm P}(t) \rF$ with a ``double 
ket'' symbol if a Floquet function is no longer regarded as an element 
of ${\mathcal H}$, but rather of the extended space 
$L_2[0,T] \otimes {\mathcal H}$. Next, one introduces the {\em quasienergy 
operators\/} 
\begin{equation}
	K^{\bm P} =  H^{\bm P}(t) + p_t \; ,
\end{equation}	
where 
\begin{equation}
	p_t = \frac{\hbar}{\ri} \frac{\rd}{\rd t} 
\end{equation}	
denotes the momentum operator conjugate to the $t$-coordinate; observe that
the periodic boundary conditions~(\ref{eq:PBC}) make sure that this operator 
is hermitian on $L_2[0,T] \otimes {\mathcal H}$ . With these conventions, the 
eigenvalue equation~(\ref{eq:EVP}) takes its proper form 
\begin{equation}
	K^{\bm P} | u_n^{\bm P}(t) \rF = 
	\varepsilon_n^{\bm P} | u_n^{\bm P}(t) \rF \; .
\label{eq:PEV}
\end{equation}
The use of adiabatic techniques for obtaining approximate solutions to the
Schr\"odinger equation~(\ref{eq:SGL}) in terms of instantaneous Floquet 
states nows rests on the following observation~\cite{BreuerHolthaus89a,
BreuerHolthaus89b,DreseHolthaus99}: The instantaneous Floquet states are 
obtained by ``freezing'' the slowly moving parameters, while retaining the 
fast, periodic $t$-dependence of the operators~(\ref{eq:HPT}). Hence, we 
require a further, time-like variable~$\tau$ in order to monitor the protocol
$\bm P (\tau)$ according to which these parameters are changed, and consider 
the Schr\"odinger-like evolution equation    
\begin{equation}
	\ri\hbar\frac{\rd}{\rd \tau} | \Psi(\tau,t) \rF
	= K^{\bm P(\tau)} | \Psi(\tau,t) \rF \; ;
\label{eq:EVO}
\end{equation}
note that its ``Kamiltonian'' $ K^{\bm P(\tau)}$ remains periodic in time~$t$ 
for {\em any\/} protocol $\bm P (\tau)$. From the solutions to this evolution
equation~(\ref{eq:EVO}) one then finds the desired solutions to the actual
Schr\"odinger equation~(\ref{eq:SGL}) by restricting the ``extended''
functions $| \Psi(\tau,t) \rF$ to the diagonal, {\em i.e.\/}, by equating
$\tau$ and $t$: Requiring 
\begin{equation}
	| \psi(t) \rangle = | \Psi(\tau,t) \rF \Big|_{\tau = t} \; ,  
\end{equation}
one immediately has~\cite{BreuerHolthaus89a,BreuerHolthaus89b}
\begin{eqnarray}
	\ri\hbar\frac{\rd}{\rd t}    | \psi (t) \rangle & = &
	\ri\hbar\frac{\rd}{\rd \tau} | \Psi (\tau,t) \rF \Big|_{\tau = t} +
	\ri\hbar\frac{\rd}{\rd t}    | \Psi (\tau,t) \rF \Big|_{\tau = t}
\nonumber \\	& = &
        \left( H^{\bm P (\tau)}(t) - \ri\hbar\frac{\rd}{\rd t} \right)
	| \Psi (\tau,t) \rF \Big|_{\tau = t} 
\nonumber \\	& & 	
	+ \ri\hbar\frac{\rd}{\rd t} | \Psi (\tau,t) \rF \Big|_{\tau = t}	
\nonumber \\	& = &
	H^{\bm P (t)}(t) | \psi(t) \rangle \; ,	
	\phantom{\frac{\rd}{\rd t}} 
\label{eq:IDE}
\end{eqnarray}
having exploited Eq.~(\ref{eq:EVO}) in the second step. 

After these somewhat painstaking preparations we are now in a position to
make use of the standard quantum adiabatic theorem~\cite{BornFock28,Kato50}:
Let us stipulate that the system is initially, at $\tau = 0$, in a Floquet
state corresponding to the parameter set $\bm P (0)$, as expressed by 
\begin{equation}
	| \Psi(\tau = 0,t) \rF = | u_n^{\bm P (\tau=0)}(t) \rF \; ,
\label{eq:INC}
\end{equation}
and let us assume that the technical propositions required by the 
adiabatic theorem are met. Then the adiabatic solution to the evolution 
equation~(\ref{eq:EVO}) takes the form 
\begin{equation}
 	| \Psi(\tau,t) \rF = \exp\left( -\frac{\ri}{\hbar} \int_0^\tau \!
	\rd \tau' \, \varepsilon_n^{\bm P(\tau')}  \right) 
	\re^{\ri \gamma_n(\tau)} | u_n^{\bm P (\tau)}(t) \rF \; ,
\label{eq:ADI}
\end{equation}
where the eigenfunctions~$| u_n^{\bm P}(t) \rF$ are determined by solving 
the instantaneous eigenvalue equations~(\ref{eq:PEV}); note that these
equations~(\ref{eq:PEV}) do not fix the phases of the eigenfunctions. Thus, 
when writing down this expression~(\ref{eq:ADI}) a certain (arbitrary, but 
differentiable) choice of these phases has implicitly been made for each value 
of $\bm P$. On the other hand, the overall phase of $| \Psi(\tau,t) \rF$ is 
uniquely fixed by the requirement that this function be a solution to the 
initial-value problem posed by Eqs.~(\ref{eq:EVO}) and (\ref{eq:INC}). 
Therefore, following Berry~\cite{Berry84}, we have introduced a phase 
$\gamma_n(\tau)$ to ensure the equality of the total phase on both sides of 
Eq.~(\ref{eq:ADI}). This phase $\gamma_n(\tau)$ then has to obey the equation   
\begin{equation}
	\dot \gamma_n(\tau) = - {\rm Im} \; \lF u_n^{\bm P(\tau)} |
	\nabla_{\bm P} u_n^{\bm P(\tau)} \rF \cdot \dot{\bm P}(\tau) \; ,
\label{eq:EFP}
\end{equation}	
as is confirmed by inserting the proposed solution~(\ref{eq:ADI}) into 
Eq.~(\ref{eq:EVO}); note that the normalization 
$\lF u_n^{\bm P} | u_n^{\bm P} \rF = 1$
implies that 
$\lF u_n^{\bm P} | \nabla_{\bm P} u_n^{\bm P} \rF$
is imaginary. Finally, implementing the general philosophy implied by the 
identity~(\ref{eq:IDE}), the desired adiabatic solution to the original
Schr\"odinger equation~(\ref{eq:SGL}) reads 
\begin{equation}
 	| \psi(t) \rangle = \exp\left( -\frac{\ri}{\hbar} \int_0^t \!
	\rd t' \, \varepsilon_n^{\bm P(t')}  \right) \re^{\ri \gamma_n(t)}
	| u_n^{\bm P (t)}(t) \rangle \; ,
\label{eq:SOL}
\end{equation}
stating that, indeed, a system starting out in a Floquet state tends to
remain in the ``connected'' Floquet state if its parameters are varied
sufficiently slowly.

These rather formal considerations require two clarifications. First, there
obviously is a geometrical Berry phase if the parameters~$\bm P$ are led
along a closed contour ${\mathcal C}$: In perfect analogy to Berry's original 
work~\cite{Berry84}, Eq.~(\ref{eq:EFP}) here yields~\cite{BreuerHolthaus89b}
\begin{equation}
	\gamma_n({\mathcal C}) = - {\rm Im} \; \oint_{\mathcal C}
	\lF u_n^{\bm P} | \nabla_{\bm P} u_n^{\bm P} \rF 
	\cdot \rd {\bm P} \; ,
\end{equation}
depending only on the loop ${\mathcal C}$ itself, but not on the way it is 
traversed. As has been pointed out by Simon~\cite{Simon83}, the standard 
adiabatic theorem provides a way of transporting a system's eigenstate along
a curve in parameter space, {\em i.e.\/}, a connection; Berry's phase therefore
is an expression of the (an)holonomy associated with this connection. In the
same sense, we now obtain a connection in $L_2[0,T] \otimes {\mathcal H}$
by {\em parallel transport of Floquet states\/}, formally given by the 
requirement that the phases of the Floquet functions occurring in 
Eq.~(\ref{eq:ADI}) be chosen such that
\begin{equation}
	\lF u_n^{\bm P} | \nabla_{\bm P} u_n^{\bm P} \rF = 0 \; ,
\end{equation}	
so that one may set $\gamma_n(t) \equiv 0$; note that the assignment of 
Floquet functions to the parameters ${\bm P}$ may not be single-valued then.
   
The second, possibly more serious clarification concerns the propositions 
required for the validity of the formal ``solution''~(\ref{eq:SOL}). Namely,
the standard adiabatic theorem~\cite{BornFock28,Kato50} requires that the 
eigenvalue of the state to be transported be separated by a certain gap 
from all others; this proposition {\em cannot\/} be fulfilled in most cases 
of Floquet transport. This is due to the Brillouin zone structure of the 
quasienergy spectrum: Suppose that we have found one solution 
$|u_n^{\bm P}(t) \rF$ to the eigenvalue equation~(\ref{eq:PEV}), and define 
$\omega = 2\pi/T$. Then one has   
\begin{equation}
	K^{\bm P} | u_n^{\bm P}(t) \re^{\ri m\omega t} \rF = 
	(\varepsilon_n^{\bm P} + m\hbar\omega) | u_n^{\bm P}(t) 
	\re^{\ri m\omega t}\rF \; ,
\end{equation}
where $| u_n^{\bm P}(t) \re^{\ri m\omega t}\rF$ again is a $T$-periodic 
Floquet function if $m$ is any integer number, be it positive, zero, or 
negative. On the other hand, all these different solutions to the eigenvalue 
problem~(\ref{eq:PEV}) amount to the same Floquet state~(\ref{eq:FLS})
in the system's physical Hilbert space~${\mathcal H}$, since
\begin{eqnarray}
	& & 
	| u_n^{\bm P}(t) \re^{\ri m\omega t} \rangle 
	\exp(-\ri [\varepsilon_n^{\bm P} + m\hbar\omega] t/\hbar)
	\phantom{\sum}
\nonumber \\	& = & 
	| u_n^{\bm P}(t) \rangle \exp(-\ri\varepsilon_n^{\bm P} t/\hbar) \; .
	\phantom{\sum}
\end{eqnarray}
Therefore, for each ${\bm P}$ the quasienergy spectrum consists of identical 
Brillouin zones of width $\hbar\omega$, with each Floquet state 
$| u_n^{\bm P}(t) \rangle \exp(-\ri\varepsilon_n^{\bm P} t/\hbar)$
leaving precisely one representative of its quasienergies
$\{ \varepsilon_n + m\hbar\omega \, | \, m =0,\pm 1,\pm 2, \ldots\}$ in 
each zone. Hence, even if we assume that a periodically driven ultracold-atoms 
system actually possesses square-integrable many-body Floquet states, which 
may be enforced by a suitable confinement, its quasienergy spectrum will 
generally cover the energy axis densely, leaving no gap that could be exploited
for strictly adiabatic transport.

Because of this lack of a quasienergy gap, the question whether or not
effectively adiabatic transport of a periodically driven Bose-Einstein
condensate could actually be exploited in a laboratory experiment is far from 
trivial. In the following section we will present model calcuations which 
indicate that there may be a window of opportunity when the parameters 
are varied at speeds which, on the one hand, are so low that the usual 
non-adiabatic transitions are suppressed, while they still remain sufficiently 
high on the other hand, such that the unusual processes associated with the 
denseness of the quasienergy spectrum do not yet figure.

\section{Numerical experiments}
\label{sec:3}

We consider the two-site model of a bosonic Josephson junction defined
by the Hamiltonian~\cite{LMG65,ScottEilbeck86,MilburnEtAl97,Leggett01}
\begin{eqnarray}
	H_0 & = & -\frac{\hbar\Omega}{2} 
	\left( \ab_1\ad_2 + \ad_1\ab_2 \right) 
\nonumber \\ & & 
	+ \hbar\kappa \left( \ad_1\ad_1\ab_1\ab_1 
	+ \ad_2\ad_2\ab_2\ab_2 \right) \; ,	
\label{eq:UJJ}
\end{eqnarray}
where the operators $\ab_j$ and $\ad_j$ effectuate the annihilation and
creation of a Bose particle at the $j$th site, obeying the commutation 
relations ($j,k = 1,2$)  
\begin{equation}
	\left[ \ab_j, \ab_k \right] = 0 	\; , \quad 
	\left[ \ad_j, \ad_k \right] = 0 	\; , \quad
	\left[ \ab_j, \ad_k \right] = \delta_{jk} \; . 
\end{equation}	
Both sites are coupled by a tunneling contact with single-particle 
tunneling frequency~$\Omega$, while the particles are interacting repulsively,
with each pair of Bosons sitting on a common site contributing the amount
$2\hbar\kappa$ to the total energy. In a typical experimental 
realization~\cite{GatiOberthaler07} the scaled interaction strength
$N\kappa/\Omega$ is on the order of unity for $N = 10^3$~particles.   

We assume that this system is subjected to external driving of the
form~\cite{HolthausStenholm01,GertjerenkenHolthaus15}  
\begin{equation}
	H_1(t) = \hbar \mu(t) \sin(\omega t)
	\left( \ad_1\ab_1 - \ad_2\ab_2 \right) \; , 
\label{eq:HFT}	
\end{equation}
so that the driving amplitude $\hbar\mu(t)$ here plays the role of the
parameter $P_1(t)$ considered in the previous section; all other parameters 
will be held constant. When occupied with $N$ particles, this system lives
in a merely $(N+1)$-dimensional Hilbert space ${\mathcal H}$, and thus is
ideally suited for numerical experiments.

\begin{figure}[th!]
\begin{center}
\includegraphics[width = 0.84\linewidth]{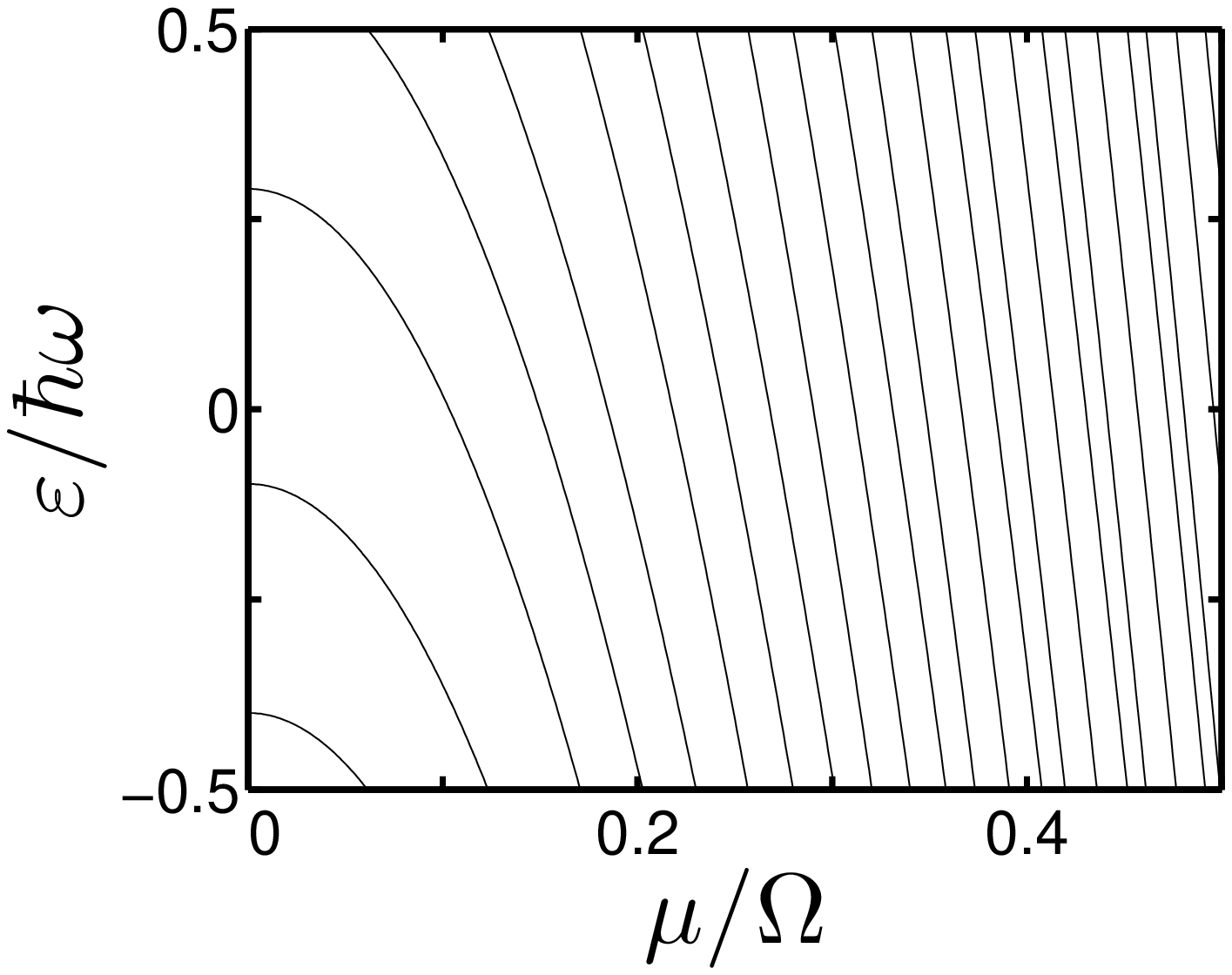}
\includegraphics[width = 0.84\linewidth]{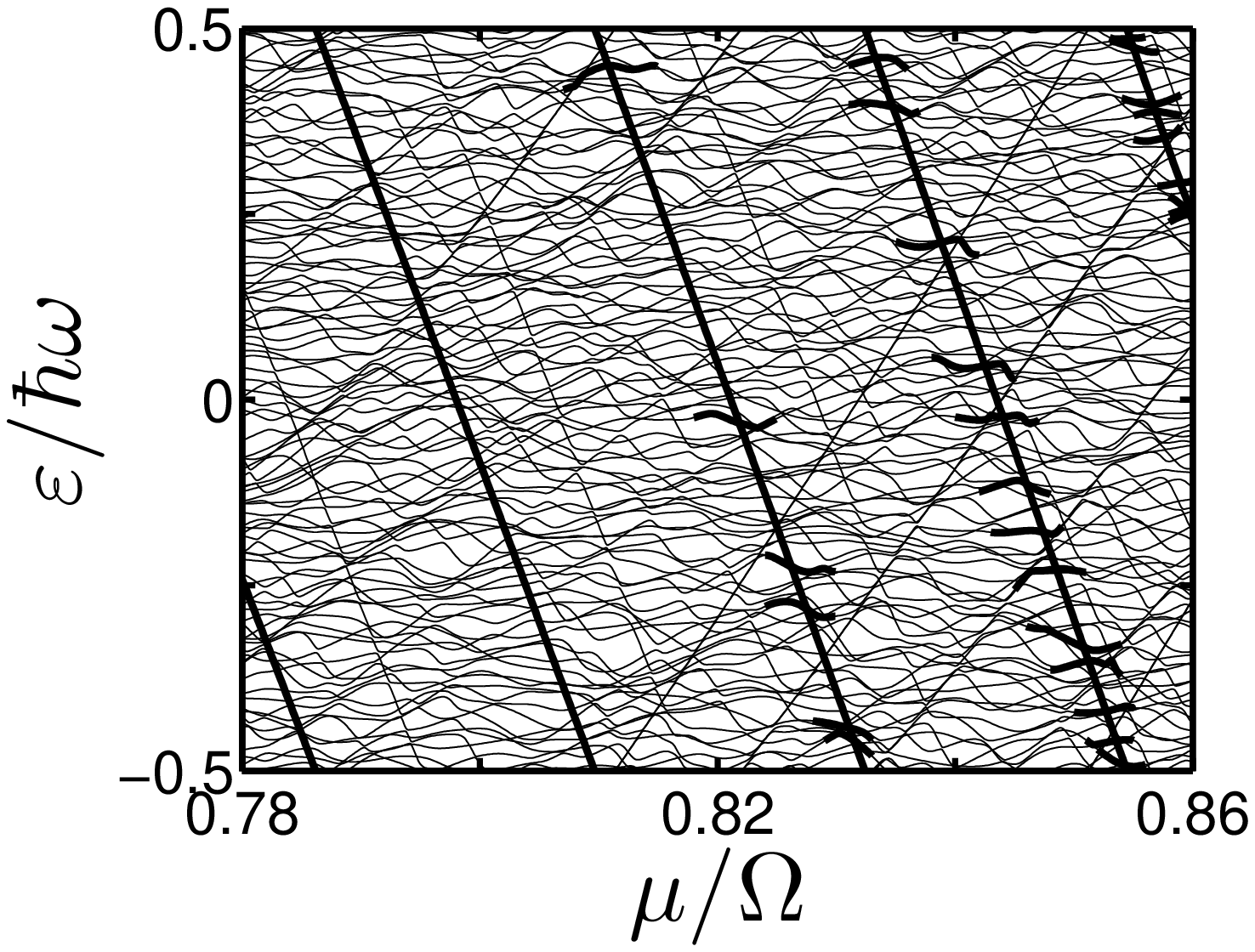}
\includegraphics[width = 0.84\linewidth]{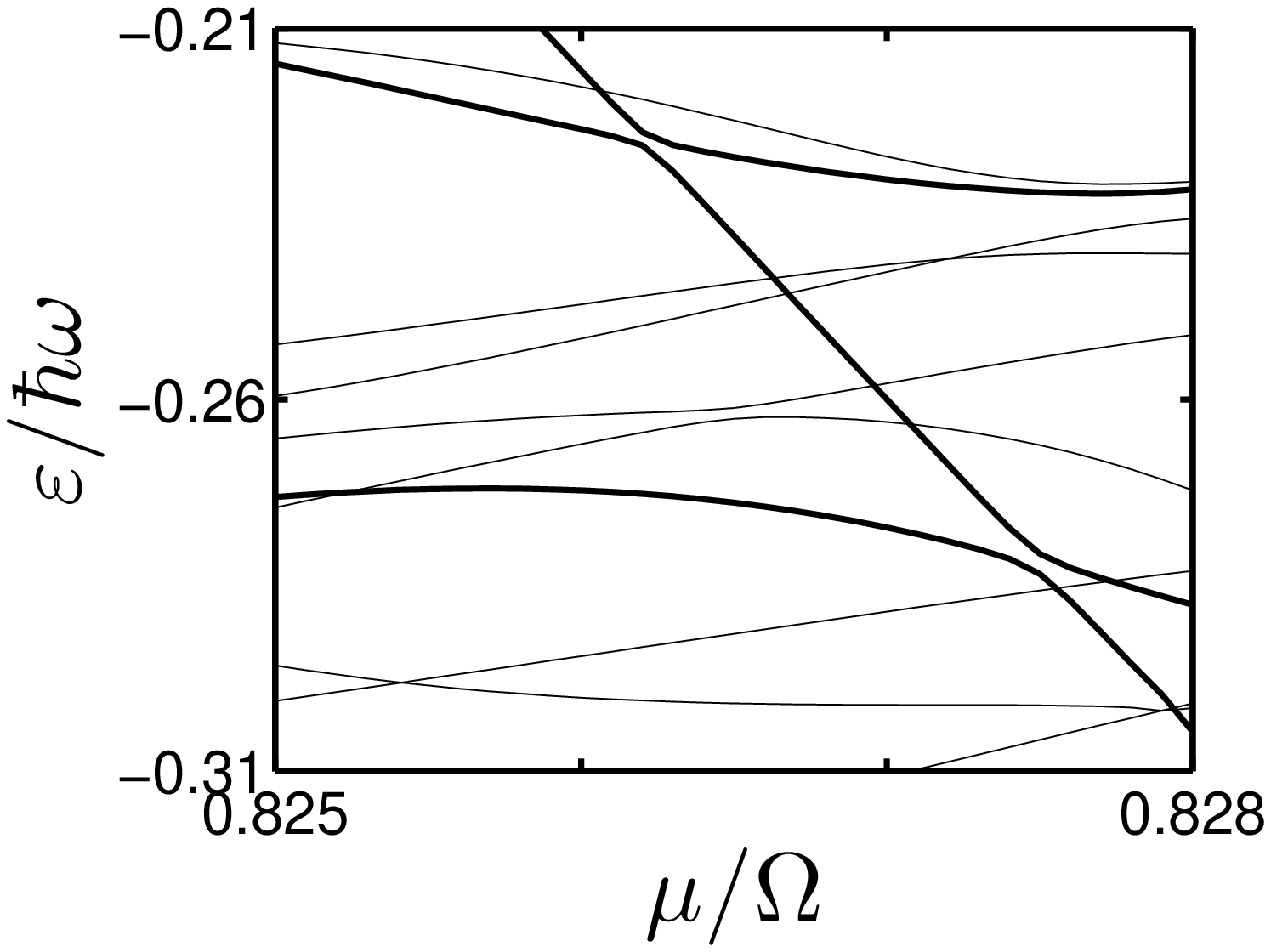}
\end{center}
\caption{(a) One Brillouin zone of quasienergy eigenvalues emerging from 
	the lowest three energy eigenstates of the undriven Josephson 
	junction~(\ref{eq:UJJ}) with $N = 100$ particles and scaled 
	interaction strength $N\kappa/\Omega = 0.95$ when subjected to 
	forcings~(\ref{eq:HFT}) with scaled driving frequency 
	$\omega/\Omega = 1.0$ and constant amplitude $\mu/\Omega$.
	On the left margin, quantum numbers~$n$ are $2$, $0$, $1$ (top to
	bottom).
	(b) Magnification of a part of the quasienergy spectrum, with all 
	eigenvalues included. The heavy line is the quasienergy emerging from 
	the ground state of the junction~(\ref{eq:UJJ}). Short heavy line 
	segments indicate avoided crossings with a gap larger than about 
	$\delta\epsilon/(\hbar\omega) = 10^{-3}$. 	  
	(c) Further magnification confirms that the quasienergy line 
	highlighted in (b) actually is broken by narrow avoided crossings. 
	These anticrossings tend to become more wide with increasing driving 
	amplitude.}
\label{F_1}	
\end{figure}

In Fig.~\ref{F_1} we display parts of the quasienergy spectrum obtained when
the system is driven with constant scaled amplitude~$\mu/\Omega$, while
$N = 100$, $N\kappa/\Omega = 0.95$, and $\omega/\Omega = 1.0$; these parameters
will be kept fixed in the following. The upper panel shows the quasienergies 
emerging from the three lowest energy eigenstates $n = 0,1,2$ of the undriven 
junction~(\ref{eq:UJJ}), reduced to the fundamental quasienergy Brillouin 
zone $-1/2 \le \varepsilon/(\hbar\omega) < +1/2$, for small scaled driving 
amplitudes $0 \le \mu/\Omega \le 0.5$. These quasienergy lines still appear 
to be smooth, providing favorable conditions for adiabatic transport. The 
middle panel then shows all 101 quasienergies of the system in the interval 
$0.78 \le \mu/\Omega \le 0.86$; the representative associated with the ground 
state $n = 0$ of the junction~(\ref{eq:UJJ}) has been highlighted. Actually, 
the corresponding Floquet state ``feels'' ({\em i.e.\/}, interacts with) the 
``background'' provided by all other states: For constant driving amplitude, 
the quasienergy operator
\begin{equation}  
	K^\mu = H_0 + \hbar\mu\sin(\omega t)
	\left( \ad_1\ab_1 - \ad_2\ab_2 \right) 
	+ \frac{\hbar}{\ri} \frac{\rd}{\rd t}
\end{equation}
remains unchanged when swapping the two sites by interchanging the indices
$1$ and $2$, and simultaneously shifting the time by half a period,
$t \to t + \pi/\omega$. The Floquet functions are even or odd under this
generalized parity, and eigenvalues belonging to the same parity should not 
cross each other, according to the von Neumann-Wigner 
theorem~\cite{NeumannWigner29}. Therefore, about half of the apparent crossings
observed in the middle panel of Fig.~\ref{F_1} actually are non-resolved 
anticrossings. This deduction is confirmed in the lower panel, which magnifies 
two of these avoided crossings. ``Relevant'' avoided crossings with a sizeable
gap only occur in the strong-driving regime; avoided crossings affecting the
Floquet state emerging from the ground state with a gap larger than about  
$\delta\varepsilon/(\hbar\omega) = 10^{-3}$ have been indicated in the 
middle panel. Here we encounter, albeit in a relatively small system only, 
a pertinent consequence of the Brillouin zone structure of the quasienergy 
spectrum: The denseness of the quasienergies gives rise to a plethora of 
avoided crossings, corresponding to multiphoton-like resonances which endanger 
adiabatic transport. However, as long as the driving amplitude remains 
sufficiently small these anti\-crossings remain possibly even undetectably 
narrow, so that the Floquet state emerging from the undriven system's 
many-body ground state is not strongly affected, and its quasienergy function 
$\varepsilon_0(\mu/\Omega)$ may be regarded as smooth and isolated in a 
coarse-grained sense~\cite{Holthaus15}.

A useful means to characterize the individual many-body Floquet states is to 
compute their respective one-particle reduced density matrix 
\begin{equation}
	\varrho_n = \left( \begin{array}{cc}
	\langle \ad_1 \ab_1 \rangle & \langle \ad_1 \ab_2 \rangle \\
	\langle \ad_2 \ab_1 \rangle & \langle \ad_2 \ab_2 \rangle 
		\end{array} \right) \; ,
\end{equation}		
and to determine the ``degrees of simplicity''~\cite{Leggett01}
\begin{equation}
	\eta_n  = 2 N^{-2} \, {\rm tr} \, \varrho_n^2 - 1 \; .
\label{eq:ETA}
\end{equation}	
Obviously $\eta_n = 1$ when $|u_n(t)\rangle$ corresponds to an $N$-fold
occupied, periodically time-dependent single-particle state, {\em i.e.\/},
to a Floquet condensate, whereas $\eta_n = 0$ when the state is maximally
fractionalized. Hence, the numerical value $0 \le \eta_n \le 1$ provides
a measure for the coherence of $|u_n(t)\rangle$.

\begin{figure}[t]
\begin{center}
\includegraphics[width = 0.84\linewidth]{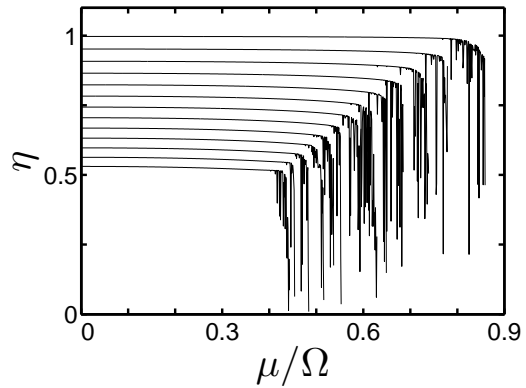}
\end{center}
\caption{Degree of coherence, as defined by Eq.~(\ref{eq:ETA}), for the 
	Floquet states connected to the lowest energy eigenstates of the 
	undriven junction~(\ref{eq:UJJ}) for $N = 100$,
	$N\kappa/\Omega = 0.95$, and $\omega/\Omega = 1.0$.  
	On the left margin, quantum numbers~$n$ are $0, 1, \ldots, 12$
	(top to bottom).}
\label{F_2}	
\end{figure}

In Fig.~\ref{F_2} we depict $\eta_n$ for $n = 0, 1, \ldots, 12$, again for 
$N = 100$, $N\kappa/\Omega = 0.95$, and $\omega/\Omega = 1.0$. Evidently 
the Floquet state emerging from the undriven ground state corresponds to 
a periodically time-dependent Bose-Einstein condensate up to roughly 
$\mu/\Omega \approx 0.8$. For higher driving amplitudes the coarse-graining 
approach mentioned above does no longer work, the web of resonances starts 
to make itself felt, and the coherence is lost.

Remarkably, here the ordering of the system's Floquet states with respect to 
their degree of coherence follows the quantum number~$n$ of the eigenstates
of the undriven junction~(\ref{eq:UJJ}). This is due to the fact that our
driving frequency~$\omega = \Omega$, times $\hbar$, is somewhat lower than 
the energy level spacing of the system~(\ref{eq:UJJ}) in the vicinity of 
its ground state; for other choices of $\omega$ the condensate-carrying 
Floquet ground state may not be connected to the unperturbed ground state
$n = 0$~\cite{GertjerenkenHolthaus14a,GertjerenkenHolthaus14b}.

Next, we explore the expected possibility of an adiabatic preparation of
a Floquet condensate: We choose a Gaussian switch-on function 
\begin{equation}
	\mu(t) = \left\{
	\begin{array}{ll}
		\mu_{\rm max} \exp\left( -\frac{t^2}{2\sigma^2} \right) 
		& , \; t \le 0 \\
		\mu_{\rm max} 
		& , \; t > 0 
	\end{array} \right.
\label{eq:SOF}	
\end{equation}		 
with steepness parameter $\sigma$. The larger the dimensionless ratio
$\sigma/T$, the longer is the time during which the system's wave function
can adjust to the changing driving amplitude.

\begin{figure}[t]
\begin{center}
\includegraphics[width = 0.84\linewidth]{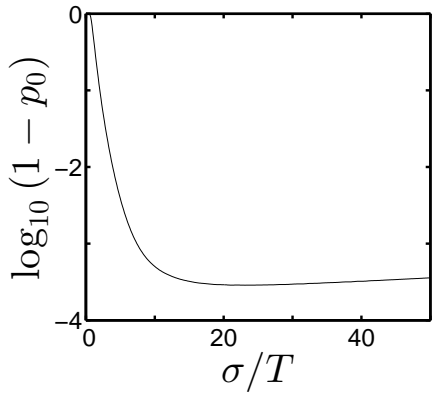}
\end{center}
\caption{Logarithm of the adiabaticity defect $1 - p_0$, with 
	$p_0 = |a_0(0)|^2$, obtained when transporting the initially prepared 
	ground state $n = 0$ of the junction~(\ref{eq:UJJ}) for $N = 100$ and 
	$N\kappa/\Omega = 0.95$  to the amplitude $\mu_{\rm max}/\Omega = 0.8$ 
	with the help of the switch-on function~(\ref{eq:SOF}), again setting 
	$\omega/\Omega = 1.0$. Observe that the best adiabaticity is achieved 
	with intermediate steepness, $\sigma/T \approx 20$. For even longer 
	turn-on times the system undergoes sizeable Landau-Zener transitions 
	at small avoided crossings.}   
\label{F_3}	
\end{figure}

We then populate the ground state $n = 0$ of the undriven junction~(\ref{eq:UJJ})
at large negative times $t_0$, when $H_1(t)$ is still negligible, and solve the 
time-dependent Schr\"odinger equation with this initial condition to obtain
$| \psi(t \rangle$ for $t_0 \le t \le 0$, employing switch-on 
functions~(\ref{eq:SOF}) with various values of the steepness~$\sigma$; 
all calculations reported in the following start at $t_0 = -5\sigma$. 
At suitable moments~$t > t_0$ the solution $| \psi(t) \rangle$ is expanded with
respect to the corresponding instantaneous Floquet states, thus obtaining their
occupation probabilities $|a_n(t)|^2$. Figure~\ref{F_3} depicts the resulting 
deviation from perfect adiabaticity at $t = 0$ when the final amplitude 
$\mu_{\rm max}/\Omega = 0.8$ has been reached: If $1 - |a_0(0)|^2$ were equal 
to zero, the initial ground state would have been transferred without loss 
into the connected Floquet state, in the sense of Eq.~(\ref{eq:SOL}). Instead, 
one observes large deviations from adiabaticity when $\sigma$ is not longer 
than a few cycles; for such sudden turn-ons the wave function simply has no 
time to adjust itself to the forcing. As expected, these deviations are 
diminished rapidly when the turn-on proceeds more slowly, being smallest 
when the steepness parameter~$\sigma$ is about $20$~cycles. However, when is 
$\sigma$ increased further, so that the steepness is {\em reduced\/} even more,
the deviations from adiabaticity {\em increase\/} again: Now the system's 
wave function does not ``jump'' more or less entirely over the small avoided 
crossings exemplified in the lower panel of Fig.~\ref{F_1}, but undergoes 
sizeable Landau-Zener transitions to the anti\-crossing Floquet 
states~\cite{BreuerHolthaus89b,DreseHolthaus99}. That is, the system becomes 
able to resolve the multitude of avoided crossings if it is given sufficient 
time. Therefore, in a system with a truly macroscopic number of particles, 
and hence with an uncomputably dense web of anticrossing quasienergies, 
an ``adiabatic limit'' in the mathematical sence, {\em i.e.\/}, for 
$\sigma/T \to \infty$, cannot be attained. Our key point is that this absence 
of a proper adiabatic limit does {\em not\/} obstruct an effectively adiabatic 
controllability for reasonably chosen, finite parameter speed, and moderate 
maximum driving amplitude.

\begin{figure}[t]
\begin{center}
\includegraphics[width = 0.84\linewidth]{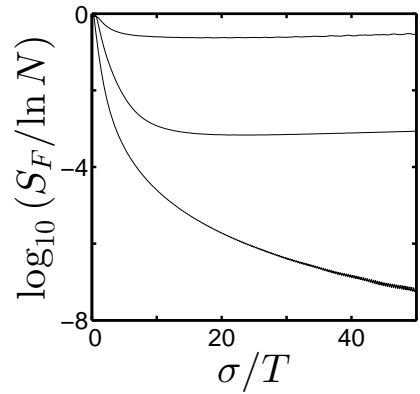}
\end{center}
\caption{Floquet entropy~(\ref{eq:FEN}), normalized to $\ln N$, resulting at 
	$t = 0$ from turn-ons~(\ref{eq:SOF}) with $\mu_{\rm max}/\Omega = 0.6$
	(lower curve), $0.8$ (middle curve), and $0.9$ (upper curve).
	Again, $N = 100$, $\omega/\Omega = 1.0$, and $N\kappa/\Omega = 0.95$.}
\label{F_4}	
\end{figure}

To substantiate this claim, we introduce the {\em Floquet entropy\/}
\begin{equation}
	S_F(t) = - \sum_n |a_n(t)|^2 \ln |a_n(t)|^2 
\label{eq:FEN}
\end{equation}  
which is zero if only one single Floquet state is populated, and takes on 
its maximum value $\ln(N+1) \approx \ln N$ when all $N+1$ Floquet states 
are populated equally. In Fig.~\ref{F_4} we show the normalized entropy 
$S_F(0)/\ln N$ resulting from turn-ons with $\mu_{\rm max}/\Omega = 0.6$, 
$0.8$, and $0.9$, respectively: As witnessed by the previous Fig.~\ref{F_2}, 
for maximum driving amplitude $\mu_{\rm max}/\Omega = 0.6$ the many-body wave
function evolving from the undriven system's ground state still remains in the 
regime where the quasienergy anticrossings cannot be resolved, so that its 
coherence is well preserved and the final entropy is still decreasing with 
increasing $\sigma$ even for ``creeping'' turn-ons with $\sigma/T \approx 50$;
of course it has to increase eventually when $\sigma/T$ is made much larger. 
The curve for $\mu_{\rm max}/\Omega = 0.8$ corresponds to the one plotted in 
Fig.~\ref{F_3}; this case falls into the parameter regime where the system 
starts to ``feel'' the anticrossings. But for $\mu_{\rm max}/\Omega = 0.9$ 
these anticrossings become so wide that the final Floquet entropy $S_F(0)$ 
already is comparable to $\ln N$, indicating that so many Landau-Zener 
transitions have taken place during the turn-on that a quite substantial 
fraction of the Floquet states has been populated to an appreciable degree, 
and the system's coherence is destroyed more or less completely.

\begin{figure}[t]
\begin{center}
\includegraphics[width = 0.84\linewidth]{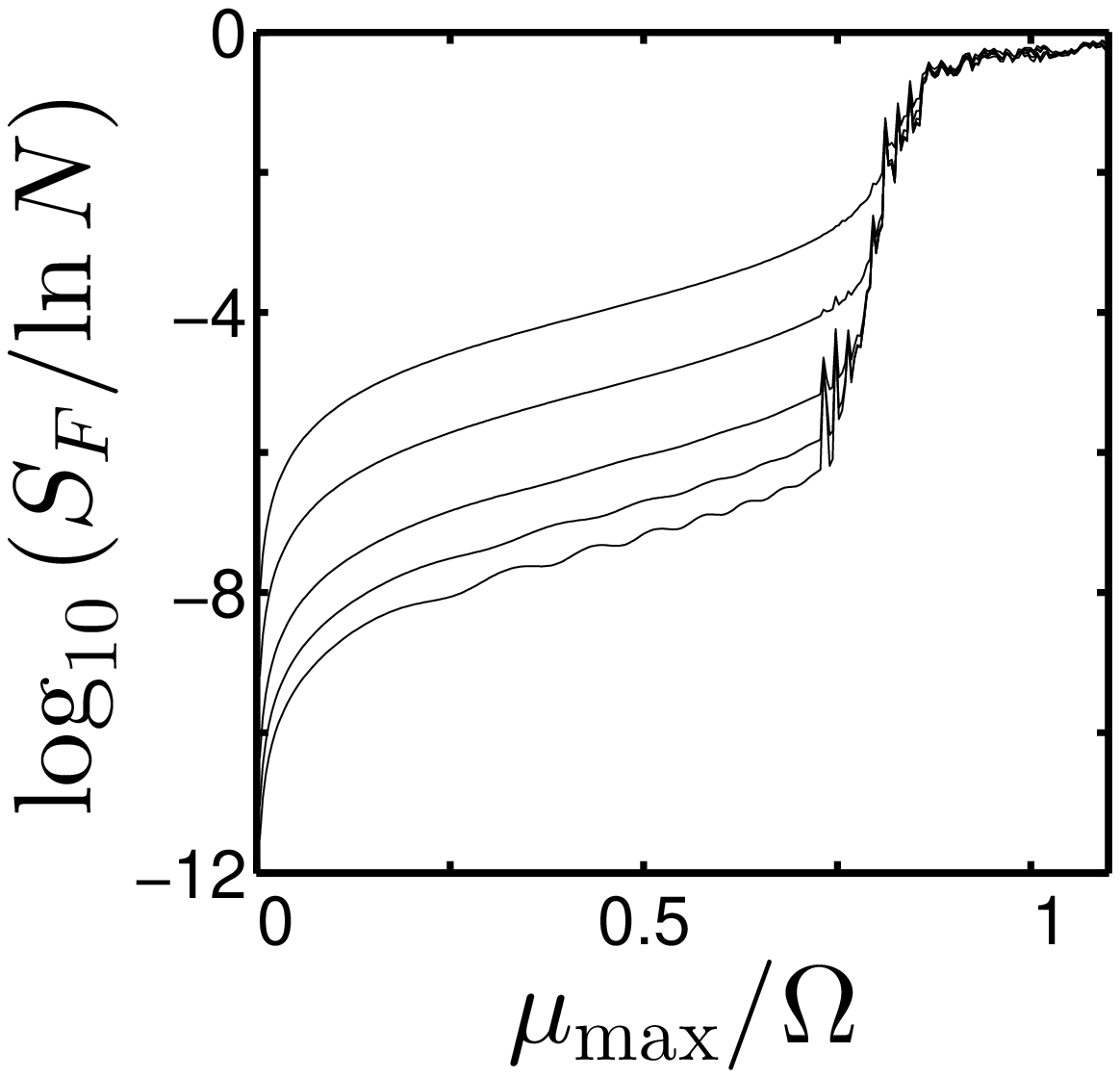}
\includegraphics[width = 0.84\linewidth]{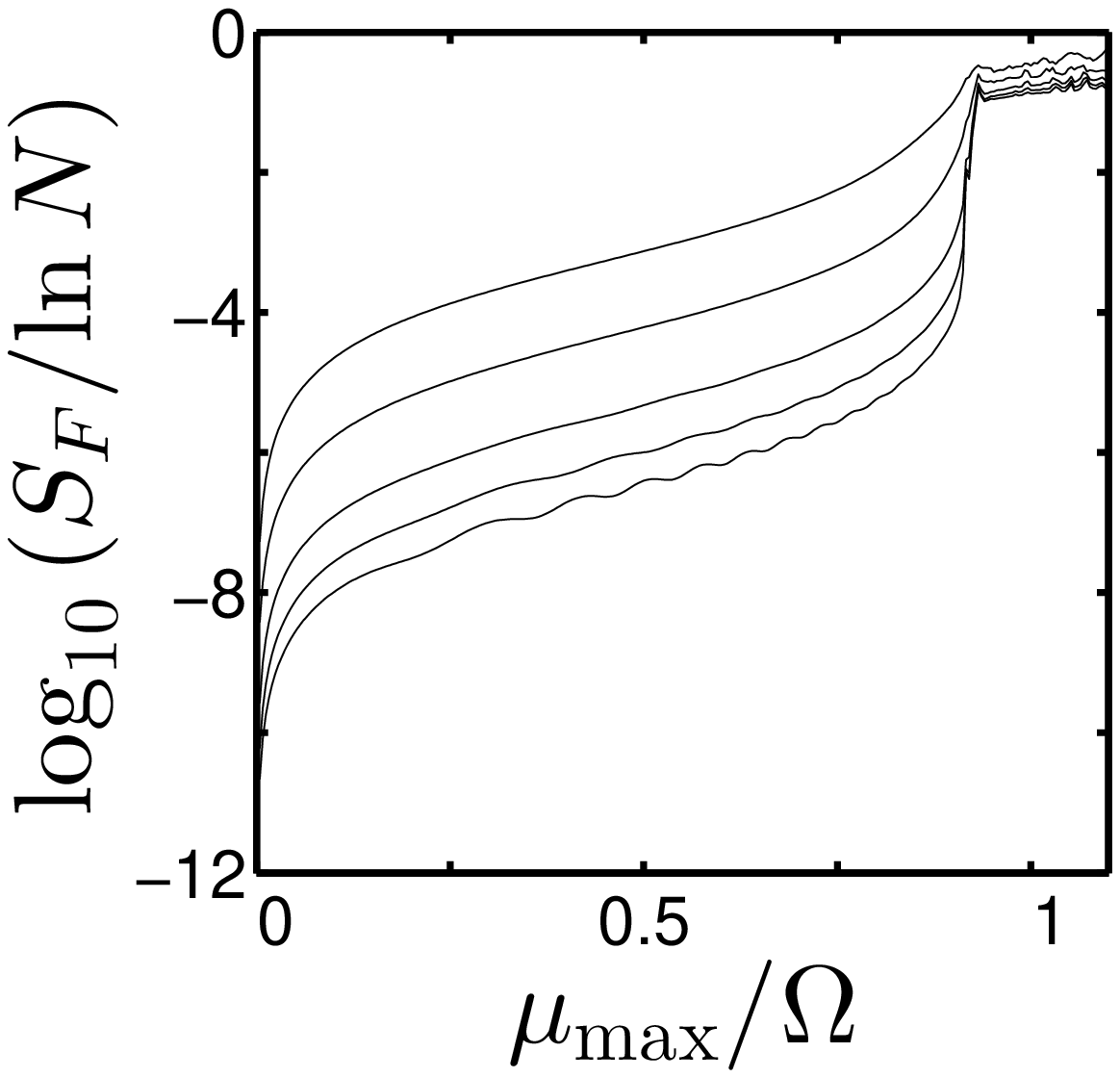}
\end{center}
\caption{Floquet entropy~(\ref{eq:FEN}), normalized to $\ln N$, resulting at 
	$t = 0$ from turn-ons~(\ref{eq:SOF}) with steepness $\sigma/T = 5$,
	$10$, $20$, $30$, $40$ (top to bottom). In the upper panel we have
	set $N = 100$, as before, whereas $N = 1000$ in the lower panel. 
	In both cases, $\omega/\Omega = 1.0$ and $N\kappa/\Omega = 0.95$.}  	
\label{F_5}	
\end{figure}

These three regimes of response --- the {\em effectively adiabatic regime\/}
in which the final Floquet entropy $S_F(0)$ is small, and decreases with
increasing~$\sigma$; the {\em transition regime\/}; and the {\em chaotic
regime\/} in which $S_F(0)$ is large, and almost independent of $\sigma$ ---
are clearly discernible in Fig~5: Here we show $S_F(0)/\ln N$ as function
of $\mu_{\rm max}/\Omega$ for several steepnesses $\sigma$. The upper panel, 
computed again for merely $N = 100$ particles, allows one to identify the 
transition regime $0.75 < \mu_{\rm max}/\Omega < 0.85$; the lower panel,
obtained for $N = 1000$, reveals a much sharper transition at 
$\mu_{\rm max}/\Omega \approx 0.85$. The fact that the main features remain
unchanged when going from $N = 100$ to $N = 1000$, while keeping 
$N\kappa/\Omega = 0.95$ at a constant value, is not trivial, because there 
are two opposing tendencies~\cite{GertjerenkenHolthaus14b}: On the one hand,
the quasienergy density in the Brillouin zone increases with~$N$, leading to 
more avoided crossings; on the other hand, the reduction of $\kappa$ with
$1/N$ implies that the individual anticrossings become more narrow. In 
combination, these trends still allow for an extended effectively adiabatic 
regime even for $N = 1000$ and larger: In this regime the $N$-particle ground 
state of the undriven Josephson junction~(\ref{eq:UJJ}) can be adiabatically 
shifted, by means of a well designed, smooth turn-on of a time-periodic driving
force, into the connected many-body Floquet state; according to Fig.~\ref{F_2}, 
this state has the coherence properties of an $N$-fold occupied, $T$-periodic 
single-particle state. In short, with judiciously chosen parameters the 
adiabatic preparation of a Floquet condensate is possible.

\section{Discussion}
\label{sec:4}

The exact numerical calculations presented in Sec.~\ref{sec:3} refer to a 
highly idealized model system. Yet, they have revealed some features which 
we believe to be generic, and which are likely to persist in actual 
laboratory set-ups not necessarily involving a condensate in a driven double 
well. We surmise the existence of windows of opportunity, {\em i.e.\/}, 
of parameter regimes enabling one to adiabatically transform a static 
Bose-Einstein condensate into a dynamic Floquet condensate without appreciable 
generation of Floquet entropy, although the denseness of the system's 
quasienergy spectrum seems to forbid a naive application of the standard 
adiabatic theorem. This prediction, which is substantiated by Fig.~\ref{F_3}, 
could be verified experimentally by subjecting a trapped Bose-Einstein 
condensate a to an oscillating drive with a smooth turn-on, and by swiching 
off the trapping potential suddenly after the maximum driving amplitude has 
been reached: Time-of-flight absorption images taken in the adiabatic regime 
then should reveal a high degree of coherence, despite the previous action of 
the possibly strong force.  

Another observation of interest, deduced from Fig.~\ref{F_5}, concerns the
possibility that the adiabatic regime may have a rather sharp border; this
is related to the recently discussed ``sudden death'' of a macroscopic wave 
function under strong forcing~\cite{GertjerenkenHolthaus15}. Here we encounter
a dynamically induced (instead of thermal) destruction of a condensate's 
coherence which can be traced to a change of the nature of the system's 
quasienergy spectrum: While the quasienergy of the state to be transported, 
when viewed as a function of the slowly changing para\-meter, is 
effectively smooth (that is, broken only by unresolvably narrow anticrossings) 
in the adiabatic regime, it becomes disrupted by a multitude of large avoided 
crossings in the coherence-killing chaotic regime. Thus, characteristic 
properties of spectra which have been investigated in great detail in the 
context of single-particle quantum chaos~\cite{Haake10,Stoeckmann07} may find 
further, unforeseen applications in the realm of forced many-particle systems.        
Again, the existence of such a ``chaos border'' should have a clear-cut 
experimental signature: In a sequence of measurements of the type sketched 
above, performed with subsequently enhanced maximum driving strengths,
one should observe a sharp coherence drop within a relatively small interval
of these maximum driving amplitudes.

\begin{acknowledgments}
We acknowledge support from the Deutsche Forschungsgemeinschaft (DFG) through 
grant No.\ HO 1771/6-2. The computations were performed on the HPC cluster 
HERO, located at the University of Oldenburg and funded by the DFG through
its Major Research Instrumentation Programme (INST 184/108-1 FUGG), and by 
the Ministry of Science and Culture (MWK) of the Lower Saxony State.
\end{acknowledgments}

\end{document}